\documentclass{article}

\usepackage{authblk}
\usepackage{multicol}
\usepackage{amsmath}
\usepackage{amsfonts}
\usepackage{amssymb}
\usepackage{geometry}
\usepackage{blindtext}
\usepackage{abstract}
\usepackage{lipsum}
\usepackage{afterpage}
\usepackage{sectsty}
\usepackage{graphicx}
\usepackage {xcolor}
\usepackage{float}

\sectionfont{\centering \fontsize{11}{12}\selectfont}

\graphicspath{ {images/} }

\raggedcolumns

\title{\vspace{-2cm} \fontsize{14}{10} \textbf{Exploring the Hidden Valley at MATHUSLA}}
\author[1]{Samuel Liebersbach}
\affil[1]{\textit{Department of Physics and Astronomy, University of Utah, Salt Lake City, UT 84112, USA}}
\affil[2]{\textit{University of California at Berkeley, Berkeley, CA 94720, USA}}

\author[1]{Pearl Sandick}

\author[2]{Abel Shiferaw}

\author[1]{Yue Zhao}
\date{}

\begin{document}

\maketitle
\begin{center}
\begin{minipage}{14cm}

Hidden valley models naturally predict numerous long-lived particles, the distinctive signatures of which would be compelling evidence for a hidden valley scenario. As these are typically low energy particles, they pose a challenge in terms of passing energy triggers in traditional searches at the Large Hadron Collider. The MATHUSLA experiment is specifically designed for the purpose of detecting long-lived particles. It also has the capability of detecting lower energy particles relative to ATLAS and CMS. In this paper, we assess MATHUSLA's potential for effectively probing hidden valley models. As a benchmark, we assume the hidden valley sector communicates with Standard Model sectors via a heavy vector propagator that couples to Standard Model quarks as well as hidden valley quarks.  We model the showering and hadronization in the hidden valley sector using PYTHIA, and study the detector acceptance as a function of the hidden valley meson's lifetime.  We find that MATHUSLA possesses significant capabilities to explore previously uncharted parameter space within hidden valley models. 

\end{minipage}
\end{center}

\begin{multicols*}{2}

\section{Introduction}\label{intro}
The Large Hadron Collider (LHC) is the flagship experiment for Standard Model (SM) and beyond Standard Model (BSM) physics studies, and has achieved extraordinary milestones, including the groundbreaking discovery of the Higgs boson. In most cases, the searches at the LHC primarily concentrate on very energetic final state particles generated from a primary vertex or events with substantial missing energy. Long-lived particles (LLPs) naturally appear in many well-motivated BSM models, and thus present the LHC with an exciting new way of studying BSM physics \cite{Giudice:1998bp, Barbier:2004ez, Chacko:2005pe, Burdman:2006tz, Meade:2010ji, Arvanitaki:2012ps, Arkani-Hamed:2012fhg, Craig:2015pha}. The search for new physics related to LLPs presents a subtle and intricate challenge. Since these particles may be soft and displaced from the primary vertex, the development of special search techniques is usually necessary, such as a novel trigger strategy or the development of new detectors.  Recently, much effort has been dedicated to the exploration and development of detectors that would be sensitive to LLPs, such as FASER($\nu$) \cite{Feng:2017uoz,FASER:2020gpr}, MATHUSLA \cite{Chou:2016lxi}, CODEX-b \cite{Gligorov:2017nwh}, and SHiP \cite{Alekhin:2015byh}.

FASER$\nu$ aims to explore the existence of light and weakly-interacting particles by studying their interactions in the forward region of the LHC detectors. It focuses on detecting neutrinos and other potential new particles that travel along the forward direction of the high energy collision. It was successfully installed in 2021 and operated during all of 2022. The first result was published in July 2023~\cite{FASER:2023zcr}. Among the proposed far detectors, FASER$\nu$ is the only one that has been successfully installed thus far. However, it is important to note that if the LLPs under investigation are not produced predominantly along the forward direction, the limited volume of FASER$\nu$ may render it less useful. Thus we do not consider FASER$\nu$ in this study.

Among all other proposed detectors, MATHUSLA (MAssive Timing Hodoscope for Ultra-Stable neutraL pArticles) has the largest volume and a very low trigger threshold. This detector is particularly well designed to detect LLPs produced off the beam axis in  LHC collisions. Its primary goal is to extend the sensitivity of particle searches by providing a large and versatile detector placed away from the main LHC interaction point. MATHUSLA would be capable of capturing and identifying LLPs that travel a significant distance, $\mathcal{O}(10)$ meters, before decaying, potentially offering insights into new physics phenomena. This nicely covers the parameter space we are interested in for this study, thus we focus on MATHUSLA in the following discussion.

In this paper we explore a benchmark hidden valley (HV) model \cite{Strassler:2006im} for LLP production, and estimate the minimum interaction cross section for LLP production that can be probed in MATHUSLA.  The organization of the paper is as follows: In section \ref{theory} we begin with a brief overview of the model used to guide our simulation.  In section \ref{experiment} we provide a more detailed description of the detector, followed by an explanation of how we simulate the detector's sensitivity in section \ref{sensitivity}.  In section \ref{check} we provide an order-of-magnitude check to corroborate our results.  Finally, we conclude with a general discussion that contextualizes our result within the current landscape of HV models.

\section{Theory}\label{theory}

HV models extend the SM by incorporating so-called hidden sectors \cite{Strassler:2006im}. These hidden sectors consist of particles and forces that exist in a secluded region not directly observable in our universe. The models suggest the presence of an additional heavy mediator particle, which acts as a bridge between the hidden sector and the observable sector. 

In this study, we explore a benchmark model where the hidden sector is characterized by a confining non-Abelian gauge group. The lightest hidden mesons, as a consequence of confinement, naturally behave as LLPs.  Such a sector is connected to the SM via a heavy vector boson $Z^{\prime}$, which is the gauge boson of an Abelian group $U(1)^{\prime}$. We assume that both the SM quarks and the hidden quarks are charged under this $U(1)^{\prime}$, and their interactions are characterized by the terms
$$\mathcal{L}\propto g_{Z^{\prime}} Z^{\prime}_\mu(Q_{SM}\bar{q}\gamma^\mu q+Q_{HV}\bar{q}_{HV}\gamma^\mu q_{HV})+\frac{1}{2}m_{Z^{\prime}}^2Z^{\prime 2},$$
where $g_{Z^{\prime}}$ is the $U(1)^{\prime}$ gauge coupling, and SM and HV quarks possess charges $Q_{SM}$ and $Q_{HV}$ respectively. To avoid various collider constraints, especially from dijet resonance searches \cite{ATLAS:2018qto,ATLAS:2020iwa,CMS:2022usq}, we assume that $Q_{SM}<<Q_{HV}$.

To simplify the simulation process, we assume that there is only one flavor of HV quark, so that the chiral symmetry in the hidden sector is not broken even though confinement happens. Thus the lightest HV mesons are one composite vector boson $\omega_h$ and one composite scalar $\eta_h$, whose masses are almost degenerate. This is in direct contrast with SM QCD, where pions are the lightest mesons due to their pseudo-Nambu-Goldstone boson nature.  By counting the number of degrees of freedom, we expect to have $N_{\omega_h}:N_{\eta_h}\simeq 3:1$.

In this benchmark model, a heavy $Z'$ can be produced in a collision and then decay to two HV quarks.  Such HV quarks undergo further showering and hadronization in the hidden sector, producing many HV mesons. These HV mesons have naturally long lifetimes due to their weak coupling to the SM sector, and they will eventually decay back to the SM sector far from the primary vertex. These decays can be hadronic or leptonic in principle.  MATHUSLA distinguishes between leptonic and baryonic events based on the multiplicity of the final state, while the energy of an event can be inferred from the geometry of the decay.  These events are distinguishable from the dominant cosmic ray background due to the directional information provided by the detector, and the rejection of charged particles by the plating on the exterior of the detector.  In addition, due to the fact that  MATHUSLA can distinguish various final states from the HV meson decays, it becomes realistic to perform an inclusive search for HV models, whereas several similar proposals rely on assumptions of certain decay channels \cite{LHC1,ATLAS:2015xit,ATLAS:2015oan,Pierce:2017taw,Yuan:2020eeu}.  The lifetimes of the HV mesons can vary widely depending on the chosen model parameters, such as the relevant decay channels and their couplings to the SM.  Such a quantity could be tuned in practice.  For this reason, we choose to keep the lifetime of the particle as a free parameter in our model.  One naturally expects that $\omega_h$ and $\eta_h$ have very different decay lifetime. For concreteness, we focus on $\omega_h$ in this study and assume $\eta_h$'s lifetime is much longer, thus its appearance in the final states does not lead to a signal.

The detailed modeling of the showering and hadronization processes in our benchmark model is highly non-trivial.  There has been much work exploring and utilizing different methods to simulate the dynamics of the final states in such events \cite{Sjostrand:2014zea,Han:2007ae, Si:1997rp, Webber:1994zd}.  Across all of these works, it appears that given a fixed number of HV mesons in the final state, different methods produce approximately similar kinematics \cite{Pierce:2017taw}.  With this in mind, we choose to use PYTHIA 8.2 to simulate the showering and hadronization processes. 

The details of the showering and hadronization processes rely on the dynamics of the hidden sector, such as the number of colors and flavors as well as the strength of the coupling constant. These processes can be properly modeled using tunable parameters within PYTHIA.  Changing those parameters affects the multiplicities of HV mesons in the final state, thus yielding distinct kinematics.  With the goal of exploring distinct kinematic final states, we will choose to change parameters in our simulations to achieve various final state multiplicities.  The details of these choices are described in section \ref{sensitivity}.

\section{Experimental Setup}\label{experiment}

Existing detectors struggle to detect LLPs for a variety of reasons.  Primarily, particles with lifetimes larger than the size of the detector will lead to displaced vertices, to which the detectors are not optimized.  It is possible for such particles to decay within the volume of the detector, but the detection of such events would depend on complex triggers that would need to distinguish the LLPs from the object mis-reconstruction or the QCD background created by the main collision event.  

\begin{figure*}[b!]
\centering
\includegraphics[width=80mm,scale=2]{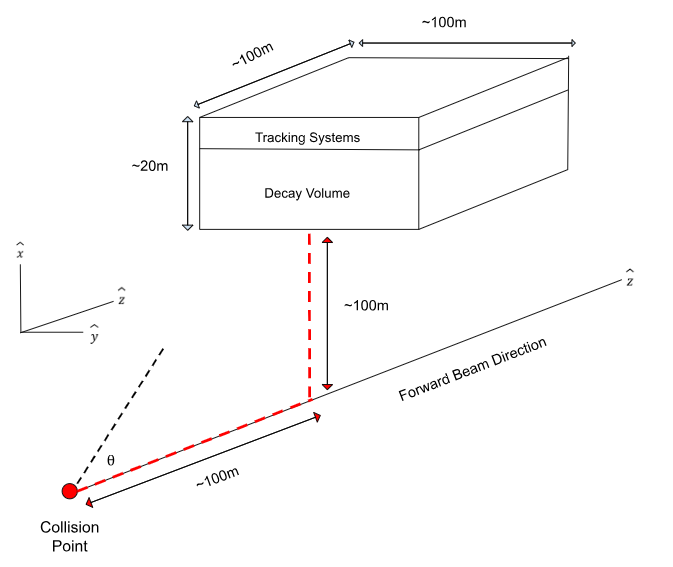}
\caption{Schematic representation of the MATHUSLA detector with associated dimensions.}
\label{fig:mathusla}
\end{figure*}

MATHUSLA allows for much greater coverage of the LLP parameter space.  This detector is proposed to be installed on the surface above either the CMS or ATLAS experiments.  As specified in~\cite{Curtin:2018mvb}, MATHUSLA's vertical and horizontal displacements from the collision point will be be roughly $100 \rm m$. The dimensions of the detector itself will be $\mathcal{O}(100 \rm m) \times \mathcal{O}(100 \rm m) \times \mathcal{O}(10 \rm m)$, which leads to a significant solid angle coverage.  Within the detector, the volume will be divided vertically into two sections.  The primary section is a 15 m tall air-filled decay region, while the secondary section consists of a 5m tall volume of tracking systems placed along the ceiling of the detector.   A schematic model of the MATHUSLA detector used in this project is provided in Fig. \ref{fig:mathusla}.

The tracking system is composed of resistive plate chambers (RPCs) which yield a timing resolution of $\sim$1ns and a spacial resolution of $\sim$1cm.  Both of these benchmarks are similar to the resolutions achieved in the muon chamber of the ATLAS detectors.  Additional technology would consist of scintillators that cover the entire exterior of the detector. The scintillators act as an early detection device for charged particles originating outside of the decay volume, and the information they take can be used to veto charged particles that may appear in the decay volume from cosmic rays or any sources not associated with the displaced vertex.  Further details of the detector can be found in~\cite{Curtin:2018mvb}.

After the production of the HV mesons, one or more of them may propagate for a macroscopic distance and decay within the primary section of the MATHUSLA detector. The decay products of the HV mesons will further propagate into the secondary section of the detector, where they will be used to reconstruct the vertex, thus creating an observable signal.  As mentioned previously, the multiplicity of the final state provides information on whether the decay was hadronic or leptonic, and the geometry of the decay provides information about the energy.  Leptonic decays usually result in both charged particles hitting the ceiling, while the number of particles hitting the ceiling is  usually much larger for hadronic decays.  The tracking system in the secondary region of the detector allows one to verify whether various decay tracks share a temporal origin within the detector, allowing further background vetos.

Since MATHUSLA is located at the surface of the Earth, one of the major concerns is the background caused by cosmic ray particles. However, cosmic rays largely propagate downward in the detector, while LLPs originate from the collision point in CMS or ATLAS.  Thus, the excellent spatial and temporal resolution of the RPCs can serve as a powerful discriminator to remove the cosmic ray background.  Given the fantastic reconstruction capability of the MATHUSLA detector, we assume zero background in the sensitivity estimation below, as the cosmic ray background can be fully subtracted for our search channel.

\section{Simulated Sensitivity}\label{sensitivity}

We now examine the sensitivity of the MATHUSLA detector with respect to our HV model.  As a benchmark, we use a $Z'$ mass of 200 GeV. Given the fact that we choose $Q_{SM}<<Q_{HV}$, dijet constraints are easily evaded \cite{ATLAS:2018qto}. Another potential constraint arises from the monojet search \cite{ATLAS:2017bfj}, given that $Z'$ predominantly decays into hidden sector particles that escape detection. Nevertheless, this constraint only imposes a relatively modest limit on the coupling between $Z'$ and SM particles, specifically, $g_{Z'} Q_{SM} \sim 0.2$ when $Z'$ has a mass of 200 GeV, which results in an upper limit of over $1000pb$ on the $Z'$ production cross-section at 14 TeV. As we will see later, MATHUSLA can achieve much lower sensitivities than this.

To demonstrate the change in sensitivity to these models due to the changing HV mass scale, we choose two benchmark HV meson masses: 0.5 GeV and 1.6 GeV.  For each of the HV meson masses we wish to explore different final state configurations of the showering and hadronization process.   In order to achieve this, we run the PYTHIA simulation with different input parameters so that the final state multiplicity of the HV mesons differs by a factor of two.  This factor of two ensures that the final state kinematics are distinct, since the final state with more HV mesons typically has a lower kinetic energy.

\begin{figure}[H]
\includegraphics[width=80mm,scale=2]{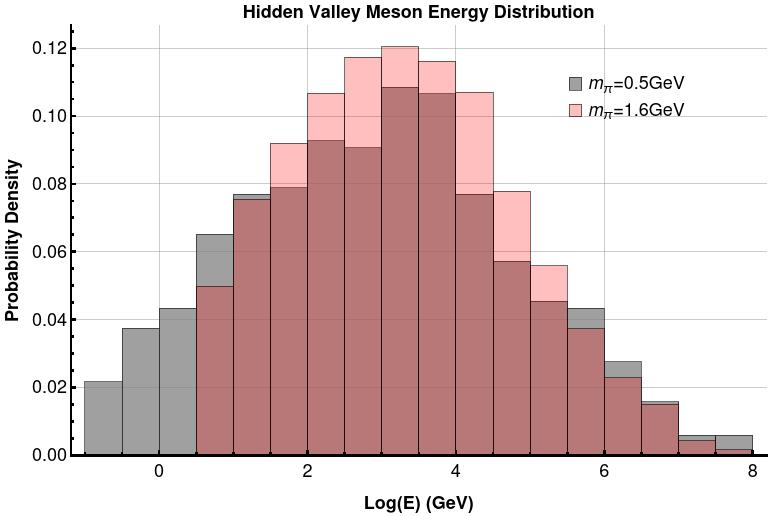}
\caption{Energy distributions for HV mesons of mass 0.5 GeV and 1.5 GeV.}
\label{fig:energy}
\end{figure}

We use PYTHIA to generate $10^{6}$ events where a $Z'$ is produced and decays to two HV quarks. PYTHIA performs showering and hadronization based on the kinematics of the HV quarks and the details of the hidden sector, yielding data for the HV mesons appearing in the final state.  For completeness, the energy distribution of these HV mesons from two of the simulations are presented in Fig.~\ref{fig:energy}.  Distributions such as these are used in subsequent calculations, while the mean boost factor for HV mesons is used in Section~\ref{check} for order of magnitude estimations.

\begin{figure}[H]
\centering
\includegraphics[width=80mm,scale=2]{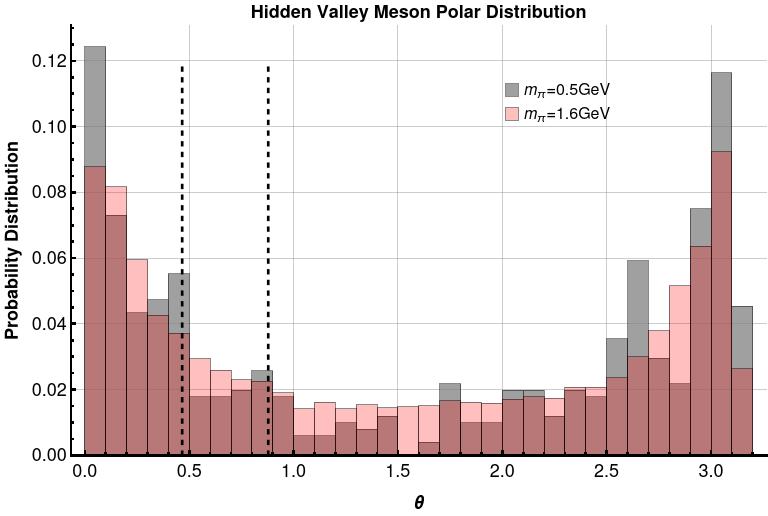}
\caption{Distribution of HV mesons in the polar angle (angle from collision axis).  We see peaks at 0 and $\pi$, corresponding to the forward and backward scattering direction respectively.  The dashed black lines represent the azimuthal angle subtended by the detector in our simulation.}
\label{fig:theta}
\end{figure}

The distribution of ejections in the polar direction is highly non-isotropic, as shown in Fig. \ref{fig:theta}.  To save computational resources, we keep only the HV mesons whose propagation directions point towards the MATHUSLA decay volume. For various choices of the HV meson proper decay lifetime, we simulate each decay following an exponential function where the exponent is scaled by the boost factor.  For each value of the lifetime, we record the number of events that have at least one HV meson decay within the primary section of the MATHUSLA detector. Dividing this number by the total number of events yields the detector acceptance $A$. Unlike many other LLP searches, there is no momentum threshold needed for one of the LLPs to propagate into the primary region of the MATHUSLA detector.

\begin{figure}[H]
\centering
\includegraphics[width=80mm,scale=2]{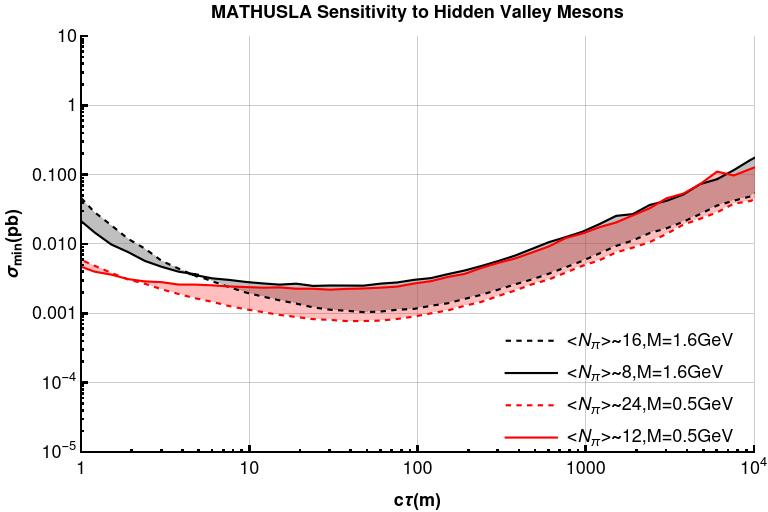}
\caption{Simulated sensitivity of the MATHUSLA detector to HV mesons produced by ATLAS.  We use masses of 0.5 GeV and 1.6 GeV as benchmarks to represent kinematically different final states.  We see that the lowest sensitivity of $10^{-3}pb$ is reached when the lifetime of the HV mesons is on the order of the separation between the production point and the decay volume.}
\label{fig:mainplot}
\end{figure}

Having calculated the detector acceptance, we are now in a position to calculate the minimum interaction cross section for each lifetime, given by $\sigma_{min}=2.7/(A \text{ }L)$ where $L$ is the luminosity, taken to be 300~fb$^{-1}$. The factor of 2.7 is a statistical factor for a 90\% C.L. assuming zero background.  The results of the simulations are shown in Fig.~\ref{fig:mainplot}.  The two bands correspond to two different HV meson masses, with the boundaries of each band defined by the multiplicity of HV mesons in the final state.  We see that  MATHUSLA achieves its best sensitivity of roughly $10^{-3}$pb when the lifetime of the particle is of the same order as the distance between the production point and the detector. We note that at lower lifetimes MATHUSLA achieves a better sensitivity if the final state contains a lower number of particles, while the opposite is true at longer lifetimes.  This is expected, as when a final state contains fewer particles, more of the energy injected into the dark sector takes the form of kinetic energy, allowing particles to live longer in the lab frame, which in turn allows them to propagate into the far detector.

It is instructive to compare our results with other searches.  There is one proposed search using the ATLAS muon detector to search for LLPs produced at the LHCb collision point \cite{Yuan:2020eeu}. Their analysis reaches a best sensitivity at $O(100)$m, which is slightly larger but comparable to the best lifetime achieved in this study. On the other hand, MATHUSLA achieves a sensitivity about three orders of magnitude better. This advantage is primarily due to the solid angle of MATHUSLA being much larger than that of the muon chamber. The momentum cutoff of ATLAS's muon chamber also hinders its sensitivity to possible soft particles that move in its direction.  

Similar benchmark scenarios have been explored in \cite{Pierce:2017taw} for LHCb, ATLAS, and CMS. The optimized lifetimes are drastically different in these searches, due to the distance of the detection regime with respect to the collision points. For lifetimes on the order of $\sim 1$m we see that MATHUSLA yields minimum cross sections an order of magnitude better than any of the aforementioned searches. They are nicely complementary to each other.  

We note that many searches mentioned above require specific decay channels of the LLPs, e.g. to muon pairs. Thanks to strong event reconstruction capabilities of MATHUSLA, we are not limited to study specific types of decays in this work. Thus one needs to properly include a correction factor caused by the decay branching ratio when comparing these results.  

Analysis similar to ours was also carried out in \cite{Chou:2016lxi} where the authors explore the detection of exotic Higgs decays by MATHUSLA. A minimum cross section of $\mathcal{O}$(fb) was achieved for lifetimes of $\sim 100$m.  This is in excellent agreement with the cross sections shown in Fig. \ref{fig:mainplot}, once one corrects for the differences of the LLP multiplicity and the assumed luminosities.

\section{Order-of-magnitude estimation}\label{check}

We provide an order-of-magnitude estimation as a self-consistency check.  The probability for a HV meson decaying in the detector can be decomposed into the product of the probability for a HV meson to propagate in the direction of the detector and the probability of the HV meson travelling an appropriate distance to decay within the detector. More explicitly, it can be written as
$$P_{detection}=P_{direction}P_{distance}.$$
In order to estimate the probability of a HV meson to propagate in the correct direction, we calculate the solid angle that the detector covers.  Taking the dimensions of MATHUSLA to be $20 \rm m \times 100 \rm m \times 100 \rm m$ and the distance from the collision point to be $ 200\rm m$, we obtain $P_{direction}\sim 10^{-2}$.  Note that the solid angle is not a perfect way of describing $P_{direction}$ given that the ejected products favor a direction along the beam, making the distribution non-isotropic, as seen in Fig.~\ref{fig:theta}. We examine how much of a deviation is induced by the non-isotropic distribution along the polar direction for our benchmarks by recording the number of events with at least one HV meson whose momentum points in the direction of MATHUSLA.  Dividing this by the total number of events and the average number of mesons in the final state gives us $P_{direction}$, which falls between $6\times 10^{-3}$ and $9\times10^{-3}$ for each of the 4 benchmarks. This aligns with the naive expectation based on magnitude.

Next we need to estimate $P_{distance}$. A small nuance is that we need the HV mesons to propagate far enough to make it to the detector, but not so far that they pass right through the detector and decay in some region where there are not tracking systems to detect them.  With these considerations, we have
$$P_{distance} = e^{\frac{-D}{\gamma c \tau}}\frac{L}{\gamma c \tau}.$$ Here we have taken a Taylor expansion with the assumption that $c\tau>>L$, where $L$ is the typical size of the detector, $\tau$ is the proper decay lifetime of the HV meson, and $\gamma$ is the boost factor. 

The first exponential function represents the probability that a HV meson survives long enough to propagate a distance $D$. The second factor describes the probability for such a particle to decay within a distance of size $L$.  For our estimation, we take $D\sim 200$ m to be the distance between the collision point and the center of the detector. This is related to the decay distance in our parameter space where we expect to achieve the best sensitivity. Further $L \sim 100$ m is taken as the typical size of the MATHUSLA detector. In order to estimate the characteristic boost factors, we take the mean value derived from the distribution of $\ln{(E)}$. These boost factors are between $5.7$ and $17$, yielding probabilities between $0.16$ and $0.06$ for all benchmarks studied here.   Generally, we see $P_{distance}\sim 10^{-1}$ when $c \tau$ is on the order of $D$.

\begin{figure}[H]
\centering
\includegraphics[width=80mm,scale=2]{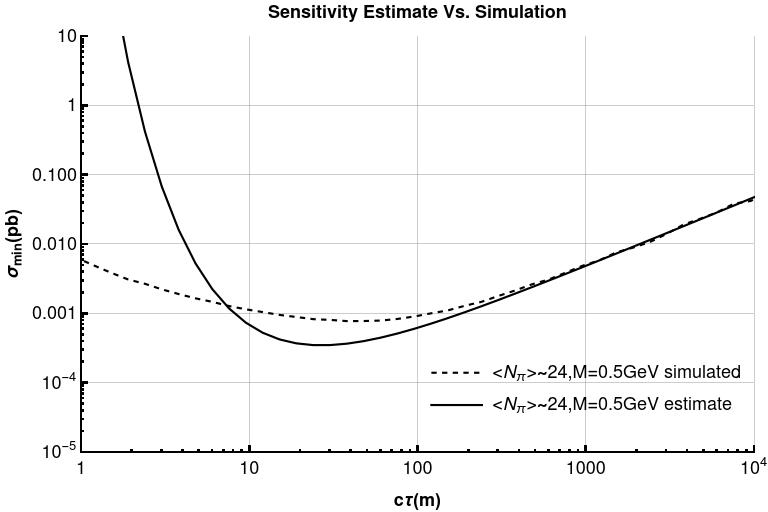}
\caption{Sensitivity comparison between the order of magnitude estimate and full simulation for a HV meson mass of 0.5GeV.  At large lifetimes we see excellent agreement.  At short lifetimes our estimate diverges since in this region our expansion assumption of $ c \tau>>L$ is no longer valid.}
\label{fig:comp}
\end{figure}

Combining our estimations gives $P_{detection}\sim 10^{-3}$.  Given this approximate detection probability, we can estimate the minimum cross section for detection.  We find that to detect $\mathcal{O} (1)$ events requires $\sigma_{min}\sim \mathcal{O}$(fb), in agreement with our results. For comparison, we show both the full-simulation result for one benchmark as well as the estimated one in Fig.~\ref{fig:comp}.  We see that the estimation generally agrees very well for large lifetimes.  For shorter lifetimes, the assumption that $c \tau>>L$ breaks down, causing a divergence in our estimate.

\section{Conclusion}\label{conclusion}

In this paper, we focus on the HV model as a benchmark for our study and estimate the extent to which MATHUSLA can enhance the ability to detect LLPs in comparison to other existing or proposed search strategies. We find that MATHUSLA possesses a remarkable capability to probe LLPs with production cross-sections on the order of femtobarns, provided that their lifetimes fall within the range of 10 to 100 meters. The sensitivity achieved by MATHUSLA surpasses that of many other experimental searches. This exceptional sensitivity can be attributed to several key factors: First, MATHUSLA's substantial detector volume provides LLPs with ample space to decay, increasing the likelihood of their detection. Second, the facility benefits from a relatively clean background environment, which enhances the signal-to-noise ratio. And third, MATHUSLA's large open angle with respect to the primary vertex allows for a comprehensive coverage of potential LLP decay products, further enhancing its capability to capture and study these particles.

The investigation of LLPs holds great importance in particle physics research due to their potential implications for extending our understanding of fundamental particles and their interactions. Numerous theoretical models have predicted the existence of such particles, making their experimental confirmation an important task in the field. Among the various experimental efforts behind LLP searches, MATHUSLA stands out as a particularly valuable tool. 

It is important to recognize that various search strategies within the field of LLP detection cover different ranges of potential decay lifetimes. Consequently, these distinct approaches complement one another and play a vital role in achieving a comprehensive exploration of the LLP landscape. Each search strategy's optimization is essential in order to ensure that no aspect of this intricate field remains unexamined. By combining the strengths of  diverse strategies, one can cover the broad spectrum of lifetimes needed for LLP searches.

\section*{Acknowledgments}
P.S.~is supported in part by NSF grant PHY-2014075.
The work of A.S.~was supported by NSF award No. 1950409.
S.L. and Y.Z. are supported by the U.S. Department of Energy under Award No. DESC0009959.

\bibliographystyle{unsrt}
\bibliography{bib}

\end{multicols*}

\end{document}